\documentclass[preprint,pre,aps]{revtex4}
\usepackage[utf8x]{inputenc}
\usepackage{graphicx}
\usepackage{epstopdf}
\usepackage{amsmath}
\usepackage{amsfonts}
\begin{document}
\title{Simple one-dimensional lattice model for lipids in water.}
\author{  Alina Ciach and Jakub P\c ekalski}
%\begin{document}
\address{Institute of Physical Chemistry,
Polish Academy of Sciences, 01-224 Warszawa, Poland}

%%%%%%%%%%%%%%%%%%%%%%%%%%%%%%%%%%%%%%%%
\begin{abstract}
A lattice model for binary mixture of lipids and water is introduced and investigated. 
  The orientational degrees of freedom of the amphiphilic molecules are taken into account 
  in the same way as in the model
  for oil-water-surfactant mixtures introduced earlier by Johan H\o ye and co-authors.
  The ground state of the model is discussed in detail, and the mean-field stability analysis 
  of the disordered phase is performed. 
  The model is compared to the recently introduced lattice model for colloidal self-assembly.
\end{abstract}
%%%%%%%%%%%%%%%%%%%%%%%%%%%%%%%%%%%%%%%%

\maketitle

%-------------------
\section{Introduction}
%---------
In 1987 Johan S. H\o ye, George Stell and Alina Ciach introduced a lattice model for ternary water, 
oil and surfactant mixtures (CHS model)\cite{ciach:88:0}. The purpose of the highly simplified,
generic model was a description of the universal properties of self-assembly in mixtures with 
amphiphilic molecules. Universality in this context means that the same properties can be found 
in many systems that differ by molecular details, but share the same property - namely the 
interaction between the amphiphilic molecules and the polar and nonpolar  molecules depends
significantly on the orientation of the former. Moreover, opposite orientations of the
amphiphiles are favorable in the neighborhood of the polar and nonpolar particles,
by which the presence of the oriented amphiphiles 
 between water and oil is energetically favored. Elimination of unnecessary details allows
for studies of the origin of such properties as the ultra-low surface tension and formation of 
lyotropic liquid crystals and bicontinuous microemulsion.

In the model introduced by Johan H\o ye and co-authors the orientational degrees of freedom of 
the water and oil molecules are disregarded, and their positions are discretized. In contrast, 
the orientational degrees of freedom of amphiphiles are taken into account, and both, the positional
and the orientational degrees of freedom of the  amphiphilic surfactant molecules are  discretized.
In a d-dimensional system 2d orientations of the amphiphiles are distinguished. 
Different orientations of the amphiphiles are treated as different components of the 2+2d component
mixture, with equal values of the chemical potential associated with each component representing 
the amphiphilic molecules. Because of the distinguished orientations of the amphiphilies,
the significantly different interactions with the head and the tail of these molecules can
be taken into account. However, the interaction between the oriented amphiphile at $x$ and 
the water molecule at $x+1$ must be significantly different than the interaction between the same
molecules, but with the water at $x$ and the amphiphile in the same orientation at $x+1$.  

Ground state of the model as well as  the stability analysis and the phase diagram obtained in 
the mean-field (MF) approximation~\cite{ciach:89:0,ciach:89:1,ciach:90:0} reproduce the key properties 
of the oil-water-surfactant 
systems in the case of vanishing spontaneous curvature of the surfactant monolayer
\cite{ciach:01:2}. For weak surfactants coexistence of the oil- and water rich phases was obtained,
whereas in the case of strong interactions between the water molecules and
the polar head of the surfactant, lamellar and bicontinuous microemulsion phases were  found to be 
stable in certain conditions~\cite{ciach:89:0,ciach:89:1,ciach:90:0}. In addition, a cubic 
bicontinuous phase with diamond symmetry can be stable \cite{ciach:91:0,ciach:92:1}. 
The surface tension between the water and the periodic phase was found to be very 
low~\cite{ciach:90:0}. The origin of the low surface tension was attributed to the amphiphilic 
nature of the surfactant. When the highly unfavorable neighborhood of the water and oil molecules 
is replaced by the water-surfactant-oil sequence of molecules, then the energetically favourable
pairs of the water and the head of the amphiphile and next the tail of the amphiphile and the oil
lead to significant decrease of the energy. An interesting observation was that at the coexistence
between the homogeneous and the periodic phases at $T=0$ the ground state is strongly degenerated,
and the surface tension vanishes~\cite{ciach:89:0}. The same property of the ground state 
was observed recently in a
model with competing short-range attractive and long-range repulsive interactions 
\cite{pekalski:13:0}. Thus, it seems that the vanishing surface tension between coexisting uniform
and periodic phases is not limited to the systems containing amphiphilic molecules.
Later the CHS model was successfully used in studies of the effects of confinement on the 
self-assembling systems \cite{ciach:02:1,babin:01:0,ciach:01:2,ciach:04:2,ciach:07:1}.

Another, even more popular model for microemulsions is based on the phenomenological Landau-Brazovskii 
functional~\cite{brazovskii:75:0} of a single scalar order-parameter (OP). 
The functional has a form of a space integral with the integrand that consists of 
a polynomial in the OP $\phi$, and in addition contains terms 
$u_2(\nabla \phi)^2$ and $u_4 (\Delta\phi)^2$ with $u_2<0$ and $u_4>0$. 
The periodic ordering is favoured by the negative $u_2$, and the length scale of 
inhomogeneities is set by the ratio $u_4/u_2$. This model was first adopted
to block copolymers by Leibler \cite{leibler:80:0}, and next to microemulsions by Gompper and Schick
\cite{gompper:94:0}. In the case of microemulsions the OP describes the local concentration 
difference between the polar and nonpolar components. However, the Landau-type functionals
 depend on a number of phenomenological parameters with no direct relation to measurable quantities.
In addition, such functionals are
not valid for strong ordering, and the formation of lyotropic liquid crystalline phases, 
where the concentration waves with 
large amplitude become stable, cannot be  reliably described.
Moreover, 
the ground states that can give valuable information on the energetically favourable ordering 
can be investigated in the lattice models much more easily than in the case of 
unrestricted positions and orientations of the molecules. Thus, the microscopic 
lattice models have many advantages.

Although the CHS model was designed for microemulsions, it could be  considered as a coarse-grained
model for symmetrical block-copolymers too. In the mapping of the CHS model on the model
for copolymers, the surfactant-occupied cell is mapped on  the part of
the copolymer containing the covalent bonding of the two types of the chain, A and B,
whereas the water- and 
oil occupied cells represent the regions occupied by the A and B monomers respectively. 
Because of the oil-water symmetry of the CHS model, the A- and B-chain lengths in the copolymer 
version of the model must be the same. 
For this reason the interesting phenomena associated with the chain asymmetry cannot be studied
in the original CHS model.  Likewise, the binary aqueous solutions of amphiphilic molecules such
as lipids could not be investigated. Only recently a continuous version of the CHS model for a
binary system was introduced \cite{ciach:11:1} in order to investigate lamellar phases in
wedge geometry.

In the present work we try to fill this gap and introduce a 1d lattice model for water and 
amphiphilic molecules inspired by the CHS model. The general version of the model allows to
study small surfactants, lipids and even 
block copolymers within the same framework. Another motivation for this study is the recently discovered 
close similarity between the self-assembly and the microsegregation observed in systems with 
competing interactions. In Ref.\cite{ciach:13:0} it was shown that the microemulsions, block copolymers,
and colloid particles interacting with competing short-range attraction and long-range repulsion
can all be described by the same Landau-Brazovskii functional. The difference between these systems
is reflected in different physical nature of the order parameter (OP). The Landau-type functionals
can be appropriate only in the case of weak ordering, i.e. for high reduced temperature. 
The question of
similarity between the amphiphilic and colloidal self-assembly in the case of strong order 
(at low $T$) remains open. 
We expect closer analogy between the assembly of the colloid particles into clusters of various
shape and assembly of amphiphiles into micelles in the binary system than in the oil-water-surfactant
mixture. The comparison between the predictions of the recently introduced model for colloidal
self-assembly and the model inspired by the CHS model should allow for a comparative study of the 
low-$T$ properties of amphiphilic and colloidal self-assembly.
We introduce the model in sec.2, describe the ground state in sec.3  and perform the MF stability 
analysis in sec.4. We close with short summary in sec.5. 

.    

%%%%%%%%%%%%%%%%%%%%%%%%%%%%%%%%%%%%%
\section{The model } 
%%%%%%%%%%%%%%%%%%%%%%%%%%%%%%%%%%%%%%
%%%%%%%%%%%%%%%%%%%%%%%%%%%%%%%%%%%%%%

\begin{figure}
\includegraphics[scale=1]{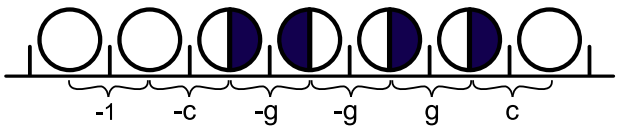}
 \caption{The interacting  pairs of occupied sites in the 1d model. 
 The open circle represents the water molecule, and 
the light and dark semicircles represent the head and the tail of the lipid molecule respectively. 
The unit of the inscripted energies is the absolute value of the  water-water interaction energy.}
\label{fig_energies}
\end{figure}

We consider a two-component mixture of water and amphiphilic molecules,  for example lipids. 
The amphiphilic molecules 
consist of a hydrophilic head and a hydrophobic tail, therefore the interactions between 
them depend on orientations. We assume that water molecules attract the
 hydrophilic head, and effectively repel the hydrophobic tail of the lipid molecule. 
 In a one-dimensional model the continuum of
 different orientations of amphiphiles is reduced to just two orientations, 
 one with the head on the left hand side and the other
 one with the head on the right hand side of the molecule. We  neglect orientational 
 degrees of freedom of the water molecules. 
In the case of charge-neutral lipids we assume nearest-neighbor interactions.
The interactions are defined in a very similar way
 as in the lattice model for ternary oil-water-surfactant systems introduced by 
 H\o ye, Stelltransactions and letters and one of  the authors in 1987
 \cite{ciach:88:0,ciach:89:0,ciach:89:1}. 
The absolute value of energy of two water molecules that occupy the nearest-neighbour sites, $-b$, 
is taken as the energy unit. We assume that the
 interaction between the water molecule and the lipid molecule in the favourable (unfavourable) 
 orientation is $-cb$ ($+cb$), and the
 interaction between two lipid molecules in the favourable and unfavourable  orientation is$-gb$ 
 and $+gb$ respectively.
The  orientations of two lipid molecules are favourable when they are oriented either head-to-head or tail-to-tail.
 The neighborhood of the polar head and the hydrophobic tail is unfavourable. 
The energies of different pairs of occupied sites are shown in Fig. \ref{fig_energies}.

We introduce the microscopic densities $\hat\rho_i(x)$ with $i=1,2,3$ denoting  water,
lipid with the head on the left
 and lipid with the head on the right respectively.
 $\hat\rho_i(x)=1$ when the site $x$ is in the state $i$ 
and $\hat\rho_i(x)=0$ otherwise. 
  Multiple occupancy of the lattice sites is excluded.
The Hamiltonian of an open system, with the chemical-potential contribution included,
can be written in the form
\begin{equation}
{\cal H}[\{\hat\rho_i\}]/b=
 \frac{1}{2}\sum_x\sum_{x'}\hat\rho_i(x)V_{ij}(x,x')\hat\rho_j(x')-\sum_x\mu_i\hat\rho_i(x),
\end{equation}
where the  summation convention for repeated indices is used,   $\mu_1b$ is the chemical potential
 of water, and since the chemical  potential of lipids,  $\mu_sb$, is independent of orientations, 
$\mu_2=\mu_3=\mu_s$. According to the above discussion of interactions,
\begin{eqnarray}
{\bf V}(x,x+1) = \left[
\begin{array}{rrr}
 -1\;& -c\;&  c\\
c\;& g\;& -g\\
-c\;& -g\;& g
\end{array}
\right]
\label{bfV}
\end{eqnarray}
and $V_{ij}(x,x-1)=V_{ji}(x,x+1)$. 

  We further restrict our attention to the liquid phase and assume close-packing. 
When each lattice site is occupied, then
the microscopic densities satisfy the condition
\begin{equation}
 \sum_{i=1}^3\hat\rho_i(x)=1.
\end{equation} 
We can eliminate $\hat\rho_1(x)=1-\hat\rho_2(x)-\hat\rho_3(x)$ in the energy contribution,
 and the Hamiltonian takes the form (up to a state-independent constant)
\begin{equation}
\label{H}
 H[\{\hat\rho_i\}]/b=\frac{1}{2}\sum_x\sum_{x'}\hat\rho_i(x)U_{ij}(x,x')\hat\rho_j(x')-\sum_x\bar\mu_1\hat\rho_1(x)
\end{equation}
where $i,j=2,3$, $\bar\mu_1=\mu_1+2-\mu_s$ and 
\begin{eqnarray}
\label{U}
{\bf U}(x,x+1) = \left[
\begin{array}{cc}
 -1+g \;\ & -1-g-2c\\
-1-g+2c\;\ & -1+g
\end{array}
\right].
\end{eqnarray}
Again, $U_{ij}(x,x-1)=U_{ji}(x,x+1)$. 
%%%%%%%%%%%%%%%%%%%%%%%%%%%%%%%%%%%%
\section{The ground state}

At $T=0$ the stable structure corresponds to the global minimum of the Hamiltonian. 
The Hamiltonian in 
Eq.(\ref{H}) for given interactions $c$ and $g$ takes the minimum for the densities
 $\hat\rho_i(x)$ that  depend on $\bar\mu$. Apart from the water-rich and lipid-rich
 phases we expect
 stability of the  periodic phase where lipid bilayers are separated by water layers.
 In the lipid-rich phase the lipids are oriented head-to-head and tail-to-tail when $g>0$. 
The values of $H$ per lattice site, $h=H/L$, in these phases are 
\begin{eqnarray}
h/b= \left\{ \begin{array}{ll}
-1-\bar\mu_1  &\quad \textrm{water}\\
-\frac{l-1+\bar\mu_1 l+2c+g(2n-1)}{l+2n} &\quad \textrm{periodic}\\
-g  &\quad \textrm{lipid}.\\
\end{array} \right.
\label{h}
\end{eqnarray} 

In the periodic phase  $l$ water occupied sites are followed by $2n$ sites occupied by
properly oriented lipid molecules.
 The coexistence lines obtained by equating $h/b$ in these phases are
\begin{eqnarray}
 \bar\mu_1=\left\{ \begin{array}{ll}
g-1 &\quad \textrm{water-lipid}\\
\frac{2c+g-3}{2} &\quad \textrm{water-periodic}\\
2(g-c) &\quad \textrm{periodic-lipid}.\\
                 \end{array}\right.
\end{eqnarray}
The three phases coexist along the triple line $g=2c-1$ and $\bar\mu_1=2(c-1)$.

Note that for the periodic phase we can write Eq.(\ref{h}) in two equivalent forms,
\begin{eqnarray}
\label{hper}
  h/b=\left\{ \begin{array}{ll}
-(1+\bar\mu_1)+\frac{2n(\bar\mu_1+1-g)+1-2c+g}{l+2n}\\
-g+\frac{l(g-1+\bar\mu_1)+g-2c+1}{l+2n}.
              \end{array}\right.
\end{eqnarray}
In the stability region of the periodic phase $h/b$ in this phase is smaller than in the other two phases,
 therefore the second terms in both expressions in (\ref{hper})  must be  negative. From the top line in
 (\ref{hper}) it follows that the lowest value of $h/b$ is assumed for $l=1$, and from the bottom line it 
follows that $n=1$ in the stable structure. However, at the coexistence with the water phase the nominator 
in the second term in the top line in 
(\ref{hper}) vanishes. Since  the periodic phase must be more stable than the pure lipid phase and therefore
 $n=1$ (see the second line in (\ref{hper})), we obtain that at the water-periodic phase coexistence the 
separation $l$ between the lipid bilayers can be arbitrary. Thus, the ground state is strongly degenerated.
 Similar degeneracy occurs at the periodic-lipid phase coexistence. In this case $l=1$ (the periodic phase 
is more stable than water). At the periodic-lipid coexistence the nominator in the second term in the bottom
 line in (\ref{hper}) vanishes, therefore the separation $2n$ between the water occupied  sites is arbitrary.

Note that the arbitrary separation between the bilayers can be interpreted as the sponge  or the disordered
 lamellar phase. On the other hand, the arbitrary separation between the bilayers is possible only at the 
coexistence between the water and the periodic phase, and can be interpreted as an arbitrary number of 
coexisting droplets of the two phases, which in addition can  have an arbitrary size. This in turn signals
 vanishing surface tension between the  periodic  and the water phases. Precisely the same properties, 
namely the coexisting droplets of the periodic and the lipid phases, and the associated  vanishing surface 
tension, are found at the coexistence between the periodic and the  lipid phases. It is interesting that in
 the 1d lattice model of particles interacting with the short-range attraction and long-range repulsion 
(SALR) potential the ground state has the same kind of degeneracy at the phase coexistence between the 
periodic and the gas or liquid phases~\cite{pekalski:13:0}.
Although the periodic phase in the case of colloidal self-assembly consists
 of periodically distributed clusters of particles which have neither shape nor interaction anisotropy, we 
obtain the same property of the ground state at the coexistence between the periodic phase and the other two phases. 

Similar degeneracy of the ground state was found earlier for the lattice model of microemulsion \cite{ciach:89:0}.
 The very low surface tension at the coexistence between the 
microemulsion and water rich phases was attributed to the amphiphilic nature of surfactant molecules. The results 
for the present model and for the model of colloidal self-assembly show that the low surface tension is a 
more general property of systems with competing interactions. When the ordered phases with  spatial inhomogeneity 
coexist with each other or with homogeneous phases, the ground state is degenerated as discussed above.
 The degenerated ground state at the phase coexistence indicates vanishing surface tension between the 
coexisting phases.
 
\begin{figure}
\includegraphics[scale=1]{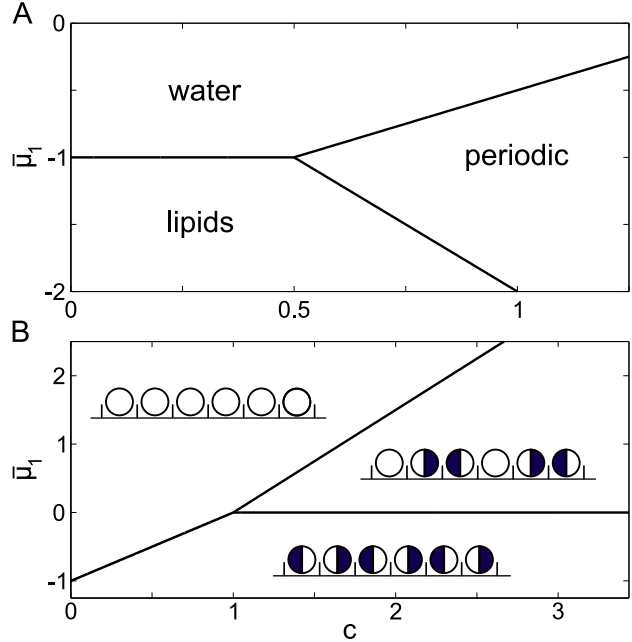}
 \caption{The ground state of the model for $g=0$ (panel A) and for $g=c$ (panel B). 
The triple point is at $(c,\bar\mu_1) = (0.5,-1)$ for $g=0$ and $(1,0)$ for $g=c$. 
On the B panel a schematic illustration of the three phases is shown in the insets inside the region of 
stability of each phase. The periodic phase is stable inside the indicated region for $l=1$ and $n=1$. 
At the coexistence with the water-rich phase the layers of water, $l$, can have an arbitrary length, 
and at the coexistence
with the lipid phase the layers of lipids, $2n$,  can have  an arbitrary length.}
 \label{fig_gs}
\end{figure}

The $(c,\bar\mu_1)$ ground state is shown in Fig. \ref{fig_gs} for two fixed values of $g$, namely for $g=0$ and for $g=c$.
 In each case the  water-lipid coexistence occurs for small values of $c$;
 $c\le 1/2$ for $g=0$ and $c\le 1$ for $g=c$. Recall that $c$ is the ratio between the energy of the pair of the 
water and  the lipidtransactions and letters molecules in the favourable orientation, and of two water molecules.
 When $c$ is larger than at the triple point (tp), than the sequence of the
 stable phases for increasing $\bar\mu_1$ is: lipids - periodic - water. All three phases coexist at the tp,
$c=1/2$, $\bar\mu_1=-1$ for $g=0$ and $c=1$, $\bar\mu_1=0$ for $g=c$. The ground states shown in Fig. \ref{fig_gs} are 
very similar to the ground state in the 1d lattice model for the SALR potential~\cite{pekalski:13:0}.
 This property confirms the observation of universality of the periodic ordering on the mesoscopic
 length scale that was derived under the assumption of weak ordering \cite{ciach:13:0}.
%%%%%%%%%%%%%%%%%%%%%%%%%%%%%%%%%%%%
\section{Mean-field stability analysis}
The grand thermodynamic potential in the mean-field (MF) approximation takes the form
\begin{eqnarray}
\label{OmegaMF}
 \beta\Omega= \frac{1}{2}\sum_x\sum_{x'}\rho_i(x)\beta U_{ij}(x,x')\rho_j(x')-\mu^*\sum_x\rho_1(x) 
+\sum_x\sum_{n=1}^3 \rho_n(x)\ln\rho_n(x)
\end{eqnarray}
where $\beta=1/(k_BT)$ and $k_B$ is the Boltzmann constant, the indices in the first term are $i,j=2,3$
 and summation convention is used for this term, $\rho_1(x)=1-\rho_2(x)-\rho_3(x)$, $U_{ij}$ is defined in Eq.(\ref{U}),
 $\mu^*=\beta\bar\mu_1$, and in thermal equilibrium the densities correspond to the minimum of $\Omega$, 
i.e. $\delta\Omega/\delta\rho_i(x)=0$ for $i=2,3$. The last term in (\ref{OmegaMF}) is the entropy contribution.
In this section we shall determine the stability of the disordered phase with respect to concentration 
waves with the wavenumber $0\le k\le \pi$. Macroscopic separation of water- and lipid-rich phases
 corresponds to $k=0$. In the crystalline  lipid phase the particles are oriented head-to-head and tail-to-tail.
 This phase  corresponds to the concentration wave with $k=\pi$, i.e. the period $\Phi$ is $2\pi/k=2$. There are 
two sublattices in this phase. The sites of the first one are numbered with even $x$, and of 
the second one with odd $x$.  
$\rho_2(x)-\bar\rho_2>0$ for $x$ belonging to one of the sublattices, whereas $\rho_3(x)-\bar\rho_3>0$ 
on the sites of  the other sublattice, where $2\bar\rho_2=2\bar\rho_3=\rho$ is the space-averaged density
 of lipids. For $0<k<\pi$  the fluctuating  bilayers of lipids oriented tail-to-tail are separated by layers of water. 
At the instability with respect to the $k$-mode 
\begin{eqnarray}
\label{det}
 \det \tilde {\bf C}(k)=0
\end{eqnarray}
where 
\begin{eqnarray}
 \tilde  C_{ij}(k)=\frac{\delta^2\beta\Omega}{\delta\tilde \rho_i(k)\delta\tilde\rho_j(-k)}
\end{eqnarray}
for $i,j=2,3$ and $\tilde \rho_i(k)=\sum_x\rho_i(x)\exp(ikx)$, with similar convention (tilde) for 
the Fourier transforms of the remaining functions.
From (\ref{OmegaMF}) and (\ref{U}) we obtain
\begin{eqnarray}
\tilde  C_{ij}(k)=\beta\tilde U_{ij}(k) +f_{ij}
\end{eqnarray}
where 
\begin{eqnarray}
 \tilde  U_{ii}(k)=2\beta^*(g-1)\cos k,\hskip1cm \tilde U_{23}(k)=\tilde U_{32}^*(k)=
-2\beta^*\big[(1+g)\cos k +2ic\sin k\big]
\end{eqnarray}
and
\begin{equation}
 f_{ij}=\Big(\frac{2}{\rho}+\frac{1}{1-\rho}
\Big)\delta^{Kr}_{ij}+\Big(\frac{1}{1-\rho}
\Big)(1-\delta^{Kr}_{ij}).
\end{equation}
From (\ref{det}) we obtain the explicit expression for the reduced temperature $T^*$ at the
instability  with respect to the density wave with the wavenumber $k$ 
for given $\rho$ 
\begin{eqnarray}
T^*(\rho,k)=-\rho\Big[P\cos k  \mp \sqrt{4c^2(1-\rho)
-Q\cos^2 k}\Big]
\end{eqnarray}
where
\begin{equation}
 P(g,\rho)=g-1+\rho
\end{equation}
\begin{equation}
 Q(c,g,\rho)=4(1-\rho)(c^2-g)-P^2(g,\rho).
\end{equation}
At the domain boundaries, i.e. for $\cos k=\pm 1$ we obtain
\begin{eqnarray}
\label{T0pi}
 T^*(\rho,k)=\left\{ \begin{array}{ll}
2\rho(1-\rho) &\quad \textrm{for $k=0$}\\
2g\rho &\quad \textrm{for $k=\pi$}.
          \end{array}\right.
\end{eqnarray}
For given $\rho$ the boundary of stability of the disordered phase corresponds to  
$T^*(\rho,k_b)$ such
 that $T^*(\rho,k)$  assumes a maximum for $k=k_b$. From the necessary condition 
 $\partial T^*(\rho,k)/\partial k=0$ 
 we obtain 
\begin{eqnarray}
\label{coskb}
 \cos^2 k_b=R,\hskip1cm R=\frac{c^2P^2(g,\rho)}{(c^2-g)Q(c,g,\rho)}.
\end{eqnarray}
The corresponding temperature is 
\begin{eqnarray}
\label{Tkb}
T^*(\rho,k_b)=4c\rho(1-\rho)\sqrt{\frac{(c^2-g)}{Q(c,g,\rho)}}.
\end{eqnarray}
 The boundary of stability of the disordered phase for given density corresponds to 
$\max(T^*(\rho,0),T^*(\rho,\pi),T^*(\rho,k_b))$.
The instability with respect to $k_b\ne 0,\pi$ can occur provided that  $0\le R\le 1$. 
The R (see Eq.(\ref{coskb})) is a positive number less than 1 for  $\rho\in (\rho_{min},\rho_{max})$,
 where the boundaries  depend on the strengths $c,g$  of the interactions. 
 For $1>\rho>\rho_{max}$ and for $0<\rho<\rho_{min}$ 
 the instability with respect to the $k_b$ mode does not occur. 
  For  $0<\rho<\rho_{min}$  the instability of the disordered phase is given by $T^*(\rho,0)$,
  and for $1>\rho>\rho_{max}$ by  $T^*(\rho,\pi)$.  
The boundaries of the interval $(\rho_{min},\rho_{max})$  are found to be 
\begin{eqnarray}
 \rho_{min}=\left\{ \begin{array}{ll}
1-\frac{g^2}{2c^2-g}  &\quad \textrm{ if   $c^2<g<2c^2$}\\                  
              1-2c^2+g    &\quad \textrm{ if  $g<c^2$ or $g>2c^2$}   
\end{array}
\right.
\end{eqnarray}
and
\begin{eqnarray}
 \rho_{max}=\left\{ \begin{array}{ll}
 1-2c^2+g &\quad \textrm{if   $c^2<g<2c^2$}\\                  
               1-\frac{g^2}{2c^2-g}   &\quad \textrm{if  $g<c^2$ or $g>2c^2$. }  
\end{array}
\right.
\end{eqnarray}
%Since the density of lipids is $0\le \rho\le 1$, the necessary condition for periodic structure is $\rho_{min}<1$ and $\rho_{max}>0$.  
We have found that $T^*(\rho,k_b)>T^*(\rho,0),T^*(\rho,\pi)$
for $\rho_{min}<\rho<\rho_{max}$ and  $0<k_b<\pi$   when $c^2>g$ and $2c^2>g+g^2$. 
   
\begin{figure}
\includegraphics[scale=1]{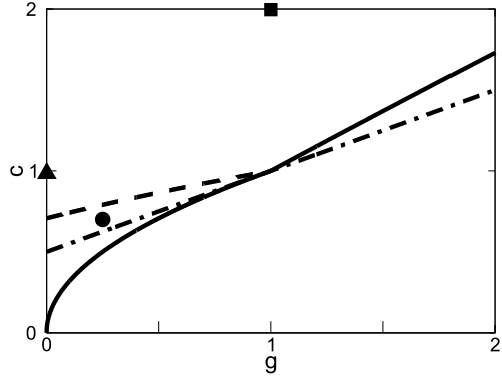}
 \caption{The plane of the interaction parameters in units of the water-water interaction 
 (see (\ref{bfV}))  with indicated regions corresponding to different types of ordering.
  Above the solid line  the disordered phase can become unstable  with respect to
 oscillatory density with the period $\Phi = 2\pi/k_b\ne 2$ for some range of densities $\rho_{min}<\rho<\rho_{max}$ 
(formation of the amphiphilic bilayers). Above the dashed line the the instability with respect to periodic ordering occurs for
$0<\rho<\rho_{max}$. Below the dashed line the separation into water-rich and lipid phases occurs for  $\rho<\rho_{min}$. 
The dash-dotted line is the projection of the three-phase line at $T=0$ on the $(c,g)$ plane. For the points marked by triangle $(1,0)$, square $(2,1)$ and circle $(c,g)=(0.7,0.25)$ the MF instability lines and the period of density oscillations along the line  $T^*(\rho,k_b)$ are presented at Figs. \ref{fig_1_0}, \ref{fig_2_1} and \ref{fig_07_025} respectively.}
\label{fig_cg}
\end{figure}
\begin{figure}
\includegraphics[scale=1]{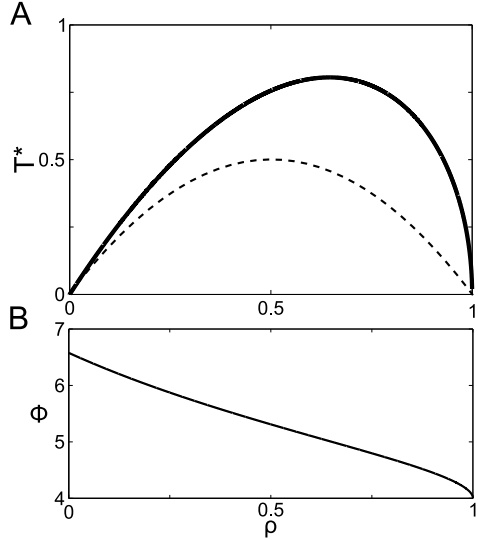}
 \caption{Panel A: The $T^*(\rho,k_b)$ (solid), $T^*(\rho,0)$ (short-dash) lines of MF instability
 with respect  to formation of the lamellar phase  (Eq.(\ref{Tkb})) and water-lipid phase separation (Eq.(\ref{T0pi})). The actual instability of the disordered phase is indicated by the thick line. Panel B: The period $\Phi$ of the density oscillations along the line of instability $T^*(\rho,k_b)$ as a function of the density of lipids. On both panels $(c,g) = (1,0)$.}
\label{fig_1_0}
 \end{figure}
\begin{figure}
\includegraphics[scale=1]{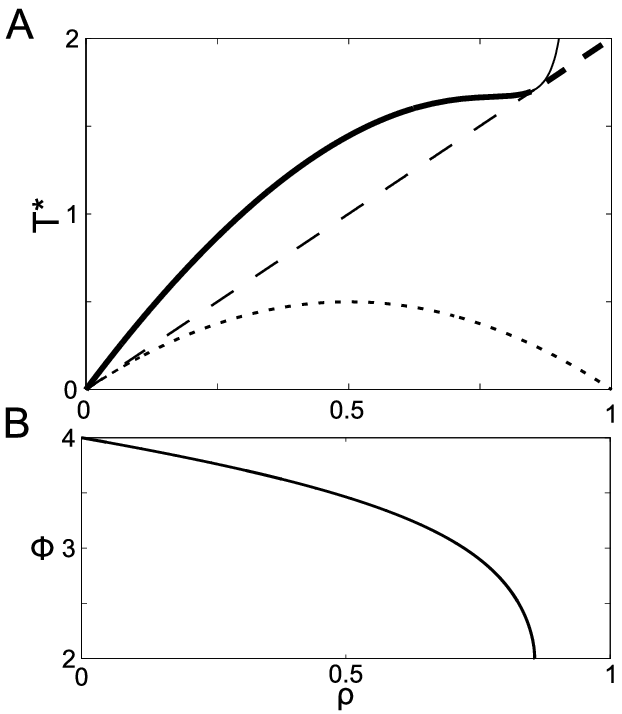}
 \caption{Panel A: The $T^*(\rho,k_b)$ (solid), $T^*(\rho,\pi)$ (long-dashed) and $T^*(\rho,0)$ (short-dash) lines of MF instability
 with respect  to formation of the lamellar phase  (Eq.(\ref{Tkb})), crystallization of lipids, and water-lipid phase separation (Eq.(\ref{T0pi})) respectively. The actual instability of the disordered phase is indicated by the thick line. Panel B: The period $\Phi$ of the density oscillations along the line of instability $T^*(\rho,k_b)$ as a function of the density of lipids. On both panels $(c,g) transactions and letters= (2,1)$.}
 \label{fig_2_1}
 \end{figure}
\begin{figure}
\includegraphics[scale=1]{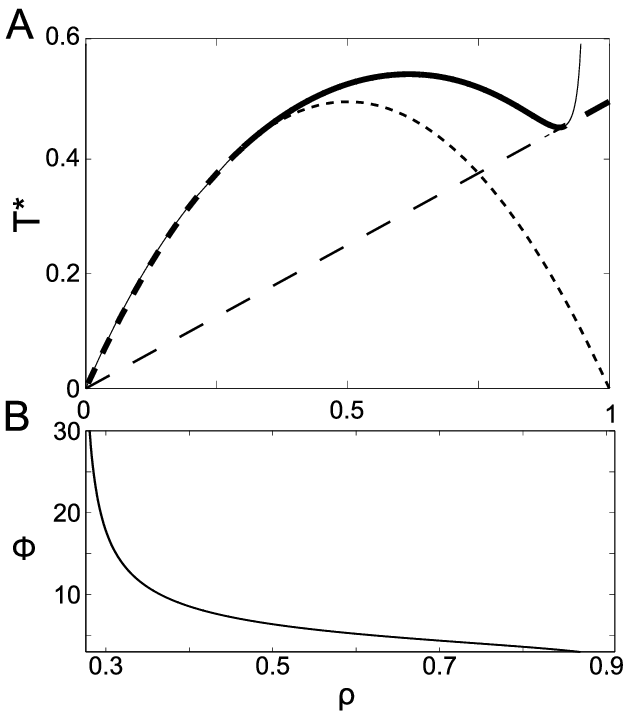}
 \caption{Panel A: The $T^*(\rho,k_b)$ (solid), $T^*(\rho,\pi)$ (long-dashed) and $T^*(\rho,0)$ (short-dash) lines of MF instability
 with respect  to formation of the lamellar phase  (Eq.(\ref{Tkb})), crystallization of lipids, and water-lipid phase separation (Eq.(\ref{T0pi})) respectively. The actual instability of the disordered phase is indicated by the thick line. Panel B: The period $\Phi$ of the density oscillations along the line of instability $T^*(\rho,k_b)$ as a function of the density of lipids. On both panels $(c,g) = (0.7,0.25)$.}
\label{fig_07_025}
 \end{figure}

The density interval $(\rho_{min},\rho_{max})$ with the above properties does not exist for $g,c$ 
located below the solid line in Fig. \ref{fig_cg}.
For such parameters the instability is with respect to water-lipid phase separation or with respect
to  ordering of lipids into a stack of oppositely oriented bilayers.

Figs. \ref{fig_1_0}A, \ref{fig_2_1}A and \ref{fig_07_025}A show the lines of instability for the parameters from different regions in Fig. \ref{fig_cg}, whereas Fig. \ref{fig_1_0}B, \ref{fig_2_1}B and \ref{fig_07_025}B show corresponding periods of density oscillations.
%%%%%%%%%%%%%%%%%%%%%%%%%%%%%%%%%%%%
\section{Discussion and summary}
We have developed a simple lattice model for water-lipid mixtures, inspired by the earlier lattice model
for miroemulsions introduced by Johan H\o ye and co-authors. Both the ground-state and the MF stability 
analysis show that the model predicts the key properties of aqueous solutions of amphiphilic molecules.
It also helps to understand the relation between the low surface tension and the degeneracy of the ground state.

We have found strong similarity of the $T=0$ phase diagrams of this model
and the lattice model for colloidal self-assembly~\cite{pekalski:13:0}.
In the model introduced in Ref.\cite{pekalski:13:0}
the nearest-neighbors attract each other,
and the third neighbors repel each other. The repulsion results from screened electrostatic 
interactions of the charged particles. The phases with oscillatory 
density or concentration occur when the repulsion in the case of colloids or attraction between 
the water and  properly oriented amphiphilic molecules in this model are strong.
When the above interations are weak, 
only two phases are present in the ground state - the gas and liquid in the SALR case and 
the water- and amphiphile rich phases in the present model. 
In both models the surface tension between
the water and the periodic phase vanishes for $T=0$, and the ground state is strongly degenerated.
Arbitrary number of arbitrarily small droplets of the coexisting phases 
can be present at the phase coexistence. This degeneracy of the ground state means that the macroscopic separation 
of the two phases at $T=0$ is not possible, since the formation of an interface does not lead to any
increase of the grand potential. At the same time, because of the microscopic size of the droplets,
one can interpret the degenerated ground state as a disordered phase. 
The region of the $T=0$ phase diagram 
corresponding to the stability of this phase is of zero measure, however, in contrast to 
the remaining, ordinary phases. The properties of this phase 
resemble the microemulsion in the present model
or a cluster fluid in the case of the model of colloids. 
Note that the inhomogeneous density, leading
to the formation of the clusters, micelles or bilayers
results from competing tendencies in the interactions, and not exclusively form the 
amphiphilic nature of the molecules.

The stability analysis shows that the main difference between the self-assembly in the amphiphilic 
and colloidal systems concerns the periodic ordering of the pure amphiphiles into the lamellar 
structure that is absent in the colloidal systems. Because of this crystal-like order of amphiphiles,
the characteristic shape of the line of instability, similar to the spinodal line of the phase
separation, is only found for $g=0$, where the amphiphiles do not order in the absence of water.
For $g>0$
the shape of the line of instability in this model is similar to the corresponding line in the
SALR system
only for low volume fraction of amphiphiles. The interesting property of the model studied in 
Ref. \cite{pekalski:13:0}, namely the phase separation for low $T$ and periodic ordering for
higher $T$ is also found in this model for relatively weak interactions between water and amphiphiles.

To conclude, the present model for water-amphiphilic systems, based on the CHS model, 
demonstrates close
similarity between different types of self-assembly. 
It also shows that there is still future for simple lattice models 
with complex phase behavior, such as the models  recently investigated by
Johan H\o ye and Enrique Lomba \cite{Hoye:08:0,Hoye:09:0,hoye:10:0}. 

%%%%%%%%%%%%%%%%%%%%%%%%%%%%%%%%%%%%
{\bf Acknowledgments} 
This work is dedicated to Prof. Johan S. H\o ye.
A part of this work  was realized within the International PhD Projects
Programme of the Foundation for Polish Science, cofinanced from
European Regional Development Fund within
Innovative Economy Operational Programme "Grants for innovation". 

\bibliography{bibliography13}

\end{document}